# Meteoritics and Cosmology Among the Aboriginal Cultures of Central Australia

Duane W. Hamacher

Department of Indigenous Studies, Macquarie University, NSW, 2109, Australia

duane.hamacher@mq.edu.au

**Abstract**

The night sky played an important role in the social structure, oral traditions, and cosmology of the Arrernte and Luritja Aboriginal cultures of Central Australia. A component of this cosmology relates to meteors, meteorites, and impact craters. This paper discusses the role of meteoritic phenomena in Arrernte and Luritja cosmology, showing not only that these groups incorporated this phenomenon in their cultural traditions, but that their oral traditions regarding the relationship between meteors, meteorites and impact structures suggests the Arrernte and Luritja understood that they are directly related.

**Note to Aboriginal and Torres Strait Islander Readers**

This paper contains the names of, and references to, people that have passed away and references the book "Nomads of the Australian Desert" by Charles P. Mountford (1976), which was banned for sale in the Northern Territory as it contained secret information about the Pitjantjatjara. No information from the Pitjantjatjara in that book is contained in this paper.

1.0     Introduction

Creation stories are the core of cosmological knowledge of cultures around the globe. To most groups of people, the origins of the land, sea, sky, flora, fauna, and people are formed by various mechanisms from deities or beings at some point in the distant past. Among the more than 400 Aboriginal language groups of Australia (Walsh, 1991) that have inhabited the continent for at least 45,000 years (O'Connell & Allen, 2004) thread strong oral traditions that describe the origins of the world, the people, and the laws and social structure on which the community is founded, commonly referred to as "The Dreaming" (Dean, 1996). This social and physical cosmology typically involves the celestial world, as deities or characters central to a cosmological story are related to the sun, moon, stars, planets, and other astronomical phenomena.

Cosmological beliefs found in the Dreamings of Aboriginal groups in Central Australia, primarily that of the Arrernte and Luritja, have an unusually strong focus on meteoritical phenomena, including meteors, meteorites, and impact craters, that are uncommon to most cultures in the world. In this paper, I provide some historical and geographical background to Aboriginal groups in Central Australia, including their astronomical traditions. I then describe the role of meteoritical phenomena and impact sites in the cosmological beliefs, ceremonies, laws, magical practices, and social structure of these groups. I finally highlight the role of these phenomena in the cosmology of other Aboriginal groups. The location of sites and towns described in this paper are given in Figure 7.

**2.0  Aboriginal Groups of Central Australia**

From 1927 to 1931, the modern-day Northern Territory (NT) was divided into North and Central Australia, the latter of which constitutes the lower half of the NT (Figure 1) and today represents one of the Territory's five regions, centered around Alice Springs. The Arrernte people are the traditional owners and inhabitants of much of this region for at least 20,000 years (Smith, 1987; Smith et al, 2001). The Arrernte, Anglicized as "Aranda" or "Arunta", are grouped into three main divisions: the Central Arrernte of Alice Springs, the Western Arrernte to the region west of Alice, and the Eastern Arrernte to the region east of Alice (Hoogenraad & Thornley, 2003). Bordering Arrernte land to the west and south is the Luritja language group, which is believed to take its name from and Arrernte word *lurinya* meaning "foreigner" (Heffernan, 1984). Traditionally hunter/gatherer peoples, both the Arrernte and Luritja relied heavily upon the landscape and skyscape for survival, the latter of which was used for time-keeping, food economics, marriage classes, and social structure (e.g. Maegraith, 1932). Close interaction between missionaries, anthropologists, and Aboriginal communities led to a great deal of literature published on Arrernte and Luritja customs and traditions, including their astronomy (see Spencer & Gillen, 1899; 1904; 1927; Strehlow, 1907, Róheim, 1945; Basedow, 1925; Maegraith, 1932; Mountford, 1976).

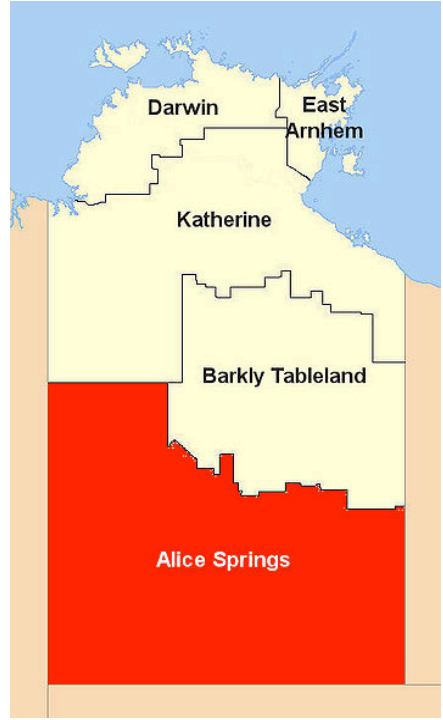

*Figure 1: The five regions of the Northern Territory, highlighting the Alice Springs region, also called "Central Australia". Image reprinted under the GNU Free Documentation License.*

**3.0     Arrernte & Luritja Astronomy**

Maegraith (1932) collected Arrernte and Luritja astronomical knowledge during the expedition of the Board of Anthropology of the University of Adelaide in August of 1929.  His informants were Arrernte and Luritja male elders from the town of Ntaria (Hermannsburg), near the border of Arrernte and Luritja country.  He interviewed each of the men separately between 8:00 and 11:00 pm, each time with a different interpreter, and only included information in the paper that were common to all of the accounts.  Because of this, summer constellations (December-March), such as Orion and Taurus, were not included.

According to Maegraith (1932), the Arrernte and Luritja partition the night sky into two equal hemispheres, or "camps", one on the western side and the other on the eastern side of the Milky Way.  The eastern hemisphere belongs to the Arrernte and the western half to the Luritja, while stars within the Milky Way incorporate both groups (see also Strehlow (1907) for a description of Arrernte and Luritja astronomical traditions).  Celestial names and stories are accounted to the various stars and celestial objects, with many stars serving as mnemonic devices or incorporated into class relationships and marriage classes.  The Arrernte and Luritja distinguished stellar colours over stellar magnitude and did not generally apply a "connect-the-dots" approach to identifying constellations as did the Greeks, instead attributing single stars or clusters to individual objects or persons in particular songs or stories (Maegraith, 1932).  Much of this sky-knowledge is restricted to initiated males, although the women have their own knowledge of the stars (Maegraith claims that the boys are told the "truth" about the stars when they are initiated). These groups acknowledged the motions of the stars, both over the night and over the year, also noting circumpolar stars (stars that never set below the horizon). While the papers by Maegraith and Strehlow reveal a treasure-trove of information regarding the sky-knowledge of Arrernte and Luritja cultures (as of 1932), it reveals little information about the role of meteors and meteorites. However, a review of the literature shows that meteors and meteorites played a significant role in the Dreamings and cosmology of these groups.

**4.0     Meteors & Meteorites**

Indigenous views of meteoritic phenomenon are varied, but meteors are generally associated with serpents, evil magic, omens of death, and punishment for breaking laws and traditions (Spencer & Gillen, 1904; Hamacher & Norris, 2010). For example, a meteor was an omen that the spirit of someone that had died far away was returning home (Basedow, 1925:296). Meteors were believed to contain an evil magic called *Arungquilta*, which was harnessed in ceremonies to inflict harm or death upon someone that broke a taboo, such as infidelity. Arungquilta could also be found in toadstools and mushrooms, which were believed to be fallen stars, and their consumption was forbidden (Spencer & Gillen, 1904). Such a taboo may have developed from the consumption of poisonous fungi that are found in the region (see Hamacher & Norris, 2010).

Two similar Arrernte rituals were performed to cause death using magical means. One ritual involved chanting a spell over a bone or stick and throwing the stick as far as possible in the direction of the intended victim. Afterwards, if the individual performing the spell saw a meteor, it was believed to represent the spirit of the person they had killed. The second ritual, recorded by Spencer & Gillen (1899:550; 1904:627; 1927:415-417) while working with the western and southwestern Arrernte, involved using a small spear-like device designed to punish a man for stealing another man's wife. The spear, endowed with Arungquilta, was thrown in the direction of the offending man's home. The evil spirit within the spear would locate the law-breaker and kill him. This form of Arungquilta was seen "streaking across the sky like a ball of fire". The men conducting the ritual would wait until a "thunderous boom" was heard, which signified that the spear had struck and killed the man. Spencer & Gillen described another form of Arungquilta that was used to punish unfaithful wives. This particular ceremony involved drawing a pictogram (Figure 2) in the dirt in a secluded area while a group of men (generally relatives of the husband) chanted a particular song. A piece of bark, representing the woman's spirit, was impaled with a series of small spears and flung in the direction they believed the woman to be, which would appear in the sky as a comet. The Arungquilta would find the woman and deprive her of her fat. After a time, the emaciated woman would die and her spirit (*ulthana*) appeared in the sky as a meteor (Spencer & Gillen, 1927:417). These descriptions seem to describe single particular events, where the rare appearance of a comet or airburst event were coincidental to the ritual and incorporated into, and explained by, the ritual itself.

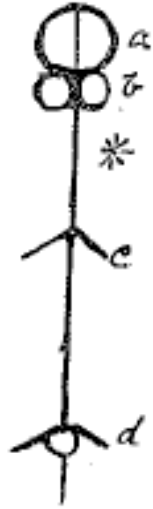

*Figure 2: From Spencer & Gillen (1927:410, Fig 126). The drawing represents a woman lying on her back, where (a) represents her head, (b) her eyes, (c) her arms, and (d) her legs. The asterix indicates where the piece of bark is placed, representing the woman's spirit.*

A decade after Spencer and Gillen's work, Strehlow (1907:30) recorded a story about meteors that did not conform to the pattern of ritual revenge/punishment. In Strehlow's account, meteors were poisonous serpents with large fiery eyes that flew through the sky and dropped into deep waterholes (see also Róheim 1945:183), while the eyes of serpents were often compared to bright stars (*ibid*), a perception shared by many Aboriginal communities in northern parts of Australia (Hamacher & Norris, 2010). This perception is illustrated by Dreamings from Ntaria (Hermannsburg), a town founded by Lutheran missionaries in 1877. One Dreaming (Róheim, 1945:184) involved a falling star that dropped into a spring where the Rainbow Serpent, Kulaia, lived (also called Kanmara by the Arrernte and Muruntu by the Luritja). In the Dreaming, a recently circumcised (initiated) boy and his brother were near the spring. The boy peered into it searching for water and was swallowed whole by Kulaia. The terrified brother ran to the camp and notified the people of this brother's demise. The boy's death caused much grief and mourning among the community, who promptly destroyed the food they had collected for the boy's initiation ceremony and left camp.

A young Arrernte woman from Ntaria told Austin-Broos (1994:142; 2009:37-38) about a star that fell to earth creating a hole on the site of the old Hermannsburg church prior to settlement of the area by Lutheran missionaries. According to the story, the people were sitting in the shade by the river. A star fell from the sky onto the spot of the old Hermannsburg church, making a large hole behind two trees. A new church was erected on that spot, with the church bell positioned between the two trees, (Austin-Broos, 1994:142). When Austin-Broos asked the young woman why the star fell on that spot, she told her it was because it was the country where Jesus had been and that it was God who sent the star to earth because it was where the "Bible mob had been". The story is the result of the heavy influence of Christian mythology introduced by the Lutheran missionaries and incorporated into the pre-existing Dreamings of the landscape, which the woman claimed were "forgotten" by the Arrernte.

In Luritja culture, meteorites were used as tools of punishment or signs of approval. Men from the Ngalia clan (of the Luritja language group) told Mountford (1976:457) that the Walanari, celestial deities who were seen as protectors of good men and punishers of bad men that lived in the Magellanic Clouds, would throw stones to the earth as punishment for breaking taboos or as signs of approval during totemic ceremonies. The Ngalia men informed Mountford that people have been killed by stones thrown by the Walanari and claimed the Walanari threw glowing stones on their camp the night after they shared sacred information with him.

While Western science proposes that amino acids, which form the basis of life, were transported to earth via comets (e.g. Elsila et al, 2009), the origin of life or humankind in Arrernte and Luritja Dreamings is also attributed to cosmic debris. A Western Arrernte account recorded by Róheim (1971:370) is that the first human couple originated from a pair of stones that were thrown from the sky by the spirit *Arbmaburinga* (or *Altierry*), a "great strong old man" that lived in a place called Jirilla to the far north. A Luritja Dreaming by Thanguwa (2008) describes how life was brought to earth by a meteorite called *Kulu*:

> *"All the animals had a big meeting. Who was going to carry the egg of life up to the universe? The Kulu was chosen. When you see where the egg of life was carried. Meteorite has landed and dropped, split three ways. This is our memory of the Kulu. And life began,"* (see Figure 3).

The Arrernte and Luritja were not the only groups in Central Australia to associate their cosmology with meteors. The Yarrungkanyi and Warlpiri people of the Northern Territory tell how Dreaming men traveled through the sky as falling stars and landed at a place called Purrparlarla, southwest of Yuendumu, bringing the Dreaming to the people (Warlukurlangu Artists, 1987:127). From this, we see that Aboriginal views of meteoritic phenomena in this region are multiple and diverse. They can be omens, tools of evil magic, progenerators of life and culture, weapons of punishment, or signs of reward. These accounts show that meteors and meteorites were an important component of the Dreamings, ritual practices, and cosmology of the Aboriginal people of Central Australia.

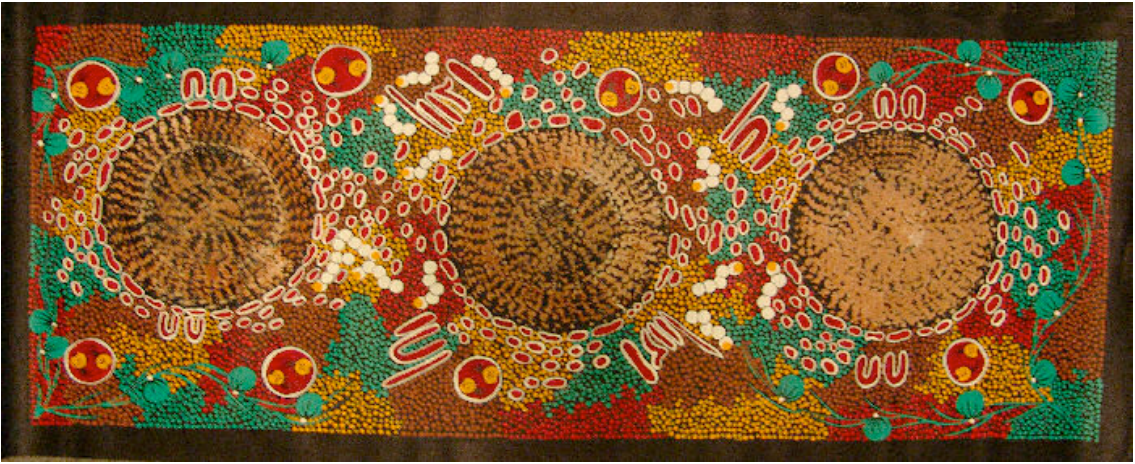

*Figure 3: A painting describing the story of Kulu, the meteorite that brought life to earth in Luritja oral traditions. Painting by Trephina Sultan Thanguwa (http://www.thanguwa.com), reproduced with permission.*

**5.0 Cosmic Impacts**

Besides the role of meteors and meteorites in Arrernte and Luritja culture and cosmology, the cosmic origin of some impact craters are acknowledged by both groups, including Henbury and Gosse's Bluff. Additionally, Wolfe Creek crater in Western Australia has an oral tradition linking it with cosmic origins.

### 5.1 Henbury Crater Field

4200±1900 years ago, a fragmented nickel-iron meteoroid struck the Central Desert, excavating thirteen craters covering an area of approximately one square kilometer (Haines, 2005). Known as the Henbury crater field (Figure 4), this event was probably witnessed first-hand by Aboriginal people and there exists enticing evidence that its memory survives in modern times. Mitchell (1934) said that older Aboriginal people would not camp within a couple of miles of the Henbury craters, referring to them as "*chindu china waru chingi yabu*", roughly translating to "*sun walk fire devil rock*". An elder Aboriginal man that accompanied Mitchell to the site explained that Aboriginal people would not drink rainwater that collected in the craters, fearing the "fire-devil" would fill them with a piece of iron. The man claimed his paternal grandfather had seen the fire-devil and that he came from the sun. Such an account suggests either a living memory of the event or recognition that the site is related to a catastrophic event in the distant past.

There are, however, conflicting accounts recorded in the literature. Brown (1975:190-191) cites the Henbury craters as an important water source to the local people while Alderman (1931:28) claims that Aboriginal people seemed to have "no interest" in the craters or any explanations regarding their origins. Although a story was recorded by Mountford (1976:259-260) describing the crater field's origin, it is attributed to an anthropomorphic figure discarding soil, forming the bowl-shape of the largest crater. Elements of this story indicate it is "women's business" (knowledge restricted to women), so I deliberately do not include the full story in this paper.

The Parks & Wildlife Commission of the Northern Territory (2002:15) cite the Arrernte name for the crater field as *Tatyeye Kepmwere* (or *Tatjakapara*) and state "*some of the mythologies for the area are known but will only be used for interpretation purposes after agreement by the Aboriginal custodians of the site*". If the site is considered sacred and secret, it may explain Alderman's claim, since his Aboriginal informants may have feigned ignorance or disinterest to prevent him from obtaining secret information. At this time, the information presented here is the only information available to the public.

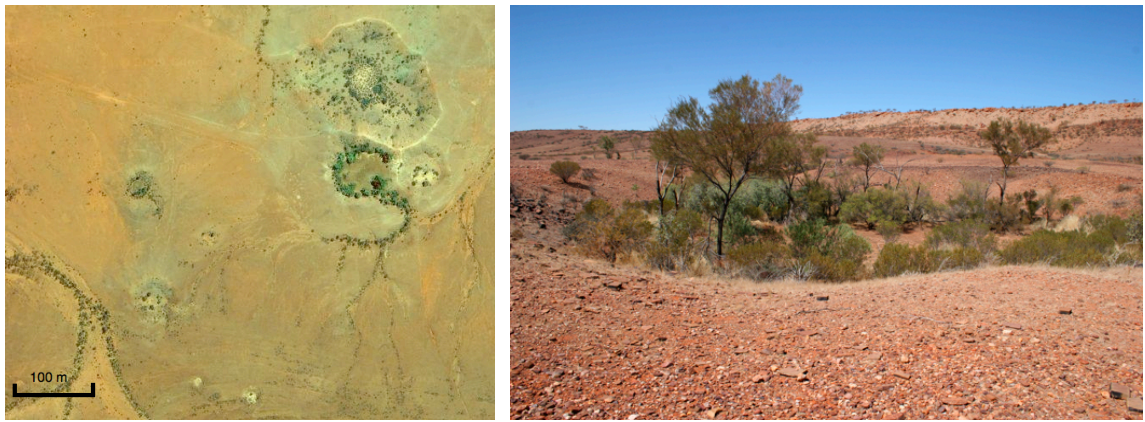

*Figure 4: Left – A satellite image of the Henbury crater filed, taken from Google Maps. Right - One of the larger Henbury craters. These craters retain water after rainfalls, creating small oases of trees and shrubs in the otherwise arid desert. The ground around these craters is still littered with iron fragments from the original impactor (photo by the author).*

### 5.2 Gosse's Bluff Crater

Towering 150 meters over the plains of the Central Desert is a 5 km wide, ring-shaped mountain range called Gosse's Bluff. This feature is the eroded central uplift of a 22 km wide complex structure that formed from a comet impact 142.5±0.8 million years ago (Milton et al, 1996). The structure, called *Tnorala* by the Western Arrernte, is registered as a sacred, but not secret, site. The Western Arrernte story of Tnorala's origin closely parallels the scientific explanation, as told by Ntaria Elder Mavis Malbunka (2009), wife of Herman Malbunka, the Traditional Owner of Tnorala (paraphrased here): in the Dreaming, a group of sky-women danced as stars in the Milky Way. One of the women grew tired and placed her baby in a wooden basket, called a turna or coolamon. As the women continued dancing, the turna fell and the baby plunged into the earth. The baby struck the earth and was covered by the turna, the force of which drove the rocks upward, forming the circular mountain range. The baby's mother, the evening star, and father, the morning star, continue to search for their baby to this day (see also Parks & Wildlife Commission of the Northern Territory 1997:1; Cauchi 2003). She continues:

> *"We tell the children don't look at the evening star or the morning star, they will make you sick because these two stars are still looking for their little baby that they lost during the dance up there in the sky, the way our women are still dancing. That coolamon, the one the baby fell out of, is still there. It shows up every winter."*

This may be a reference to the "galactic bulge" - the largest and brightest region of the Milky Way, which represents the center of our galaxy and looks similar in shape to a turna or coolamon. It is seen in the region bordering Sagittarius and Scorpius and is prominent in the winter night skies (see Figure 5). The motif of the dancing stars (women) may be attributed to the Phi Sagittariids, a meteor shower that radiates from the center of the galactic bulge between 1 June and 15 July, when the Milky Way is high in the winter night sky, although there is currently no ethnographic evidence to support this speculation.

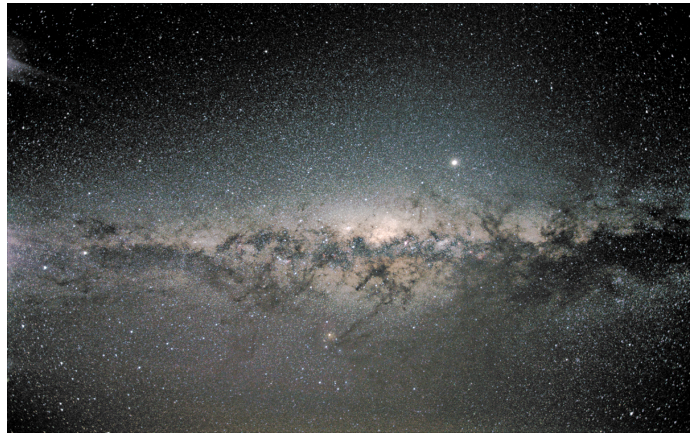

*Figure 5: The galactic bulge in the Milky Way as seen from Victoria, Australia. Image by Dr. Russell Cockman, reproduced with permission (http://www.russellsastronomy.com/)*

### 5.3 Wolfe Creek & Other Craters

Henbury and Gosse's Bluff are not the only Australian impact craters to have associated Aboriginal oral traditions. Wolfe Creek crater in northeastern Western Australia, known to the local Djaru as *Kandimalal*, is attributed to a cosmic impact, representing the spot where the Rainbow Serpent (whose eyes are seen as a meteor) crashed to the earth (Goldsmith, 2000; Sanday, 2007; see Figure 6), despite the impact having occurred some 300,000 years ago (Shoemaker et al, 2005), predating human habitation of Australia by 250,000 years.

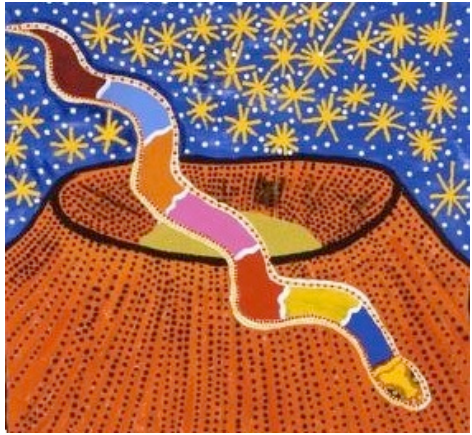

*Figure 6: Wolfe Creek Crater and the Rainbow Serpent. Painting (2000) by Boxer Milner, a Djaru Elder from Billiluna, Western Australia. Image reproduced with permission, courtesy of Peggy Reeves Sanday (2007), http://www.sas.upenn.edu/~psanday/Aboriginal/serpent4.html.*

Boxhole crater, located ~170 km northeast of Alice Springs, is a smaller impact crater (D = ~170 m) that dates to between 5,400±1,500 years using $^{14}$C exposure (Kohman & Goel, 1963) and ~30,000 years using $^{10}$Be/$^{26}$Al exposure (Shoemaker et al, 1990; 2005; Haines 2005:484-485). It is possible that this event was witnessed firsthand, but Madigan (1937:190) claimed that the local Aboriginal people seemed to have "no interest" in the crater and nothing further is reported in the literature. Madigan & Alderman (1939:355–356) claim that Aboriginal people would steer clear of the nearby Huckitta meteorite and suggested that they were in awe of the stone, perhaps considering it sacred. Liverpool crater in Arnhem Land, Northern Territory (Haines, 2005) is described by local Aboriginal people as the nest of a giant catfish (Shoemaker & MacDonald, 2005). See Hamacher & Norris (2009; 2010; 2011) for a full treatise on cosmic impacts, meteors, and comets in Aboriginal oral traditions, respectively.

7.0     **Conclusion**

A review of the literature regarding Arrernte, Luritja, and other Central Australian cultures reveals that meteoritic phenomena played a significant role in their cosmology, rituals, and oral traditions. These accounts suggest the Arrernte and Luritja recognized meteors as objects from the sky that would occasionally fall to the earth, and in some cases, cause impact craters. In some cases, the appearance of a comet or an airburst during a ritual was incorporated into those traditions, which reinforced the power of the spell and the spell-caster. Meteors and meteorites have diverse perceptions and purposes, but were generally viewed negatively as omens of death, possessing evil magic, and tools of death or punishment. In some cases, they were perceived in a positive light, as bringers of life or signs of approval. Impact structures, such as Gosse's Bluff (Tnorala), are attributed to cosmic impacts, mirroring the Western scientific explanation, while Dreamings of the Henbury craters suggest the impact was witnessed firsthand and its memory survives in modern times.

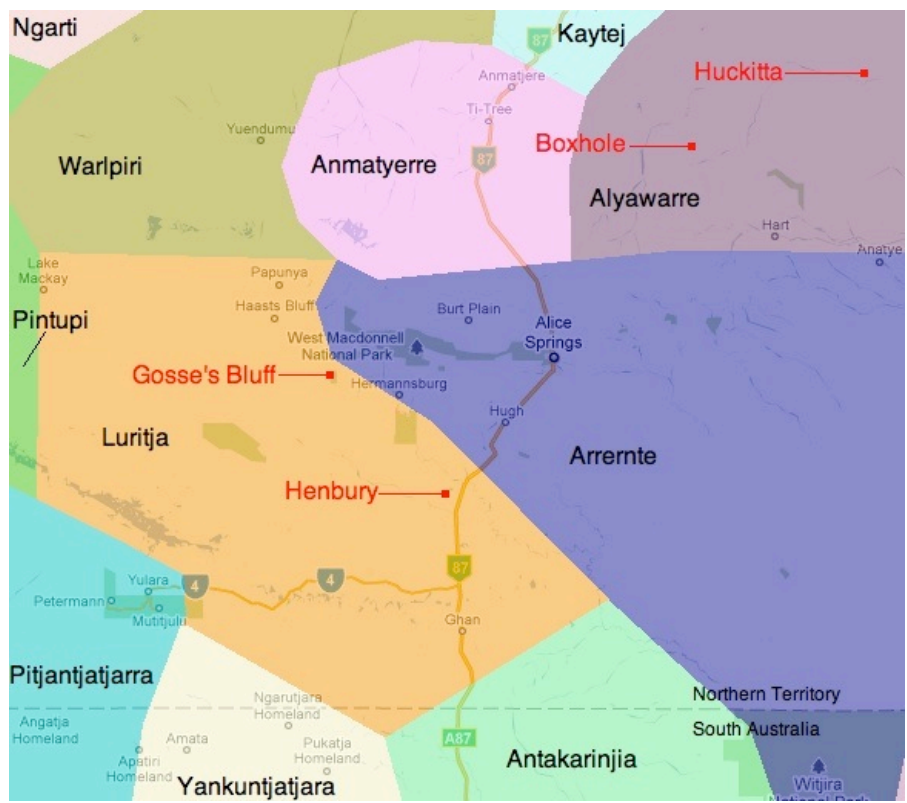

*Figure 7: Locations of impact craters, meteorite finds, and other sites described in this paper, shown in red. The coloured areas show the general boundaries of different Aboriginal language groups, labeled in black. These boundaries are not exact and serve only as an estimate. Image from Google Maps, with language group boundaries superimposed using language data © Aboriginal Studies Press, AIATSIS and Auslig/Sinclair, Knight, Merz, 1996. Created by David R Horton.*

**Acknowledgements**

I would like to thank Ray Norris and John McKim Malville. This research was funded by the Research Excellence Scholarship at Macquarie University in Sydney, Australia.

**References**

Alderman, A.R. (1931). The meteorite craters at Henbury, Central Australia with an addendum by L.J. Spencer, *Mineralogical Magazine*, 23, 19-32.

Austin-Broos, D. (1994). Narratives of the encounter at Ntaria, *Oceania*, 65, 131-150.

Austin-Broos, D. (2009) *Arrernte Present, Arrernte Past: Invasion, Violence, and Imagination in*

*Indigenous Central Australia*. Chigaco, University of Chicago Press.

Basedow, H. (1925). *The Australian Aboriginal*. Adelaide, F. W. Preece and Sons.

Brown, P.L. (1975). *Comets, Meteorites, and Men.* London, The Scientific Book Club.

Cauchi, S. (2003). Outer space meets outback and a bluff is born. *The Age* 18 June 2003, http://www.theage.com.au/articles/2003/06/17/1055828328658.html

Collins, G.S., Melosh, H.J. and Robert A.M. (2005). Earth Impact Effects Program: A Web-based computer program for calculating the regional environmental consequences of a meteoroid impact on Earth, *Meteoritics & Planetary Science*, 40(6), 817–840.

Dean, C. (1996). *The Australian Aboriginal Dreamtime: its history, cosmogenesis cosmology and ontology*. Geelong, VIC, Gamahucher Press.

Elsila, J.E.; Glavin, D.P. and Dworkin, J.P. (2009). Cometary glycine detected in samples returned by Stardust, *Meteoritics & Planetary Science*, 44(9), 1323-1330.

Haines, P.W. (2005). Impact Cratering and Distal Ejecta: The Australian Record, *Australian Journal of Earth Sciences*, 52, 481-507.

Hamacher, D.W. and Norris, R.P. (2009) Australian Aboriginal Geomythology: Eyewitness Accounts of Cosmic Impacts? *Archaeoastronomy*, 22, 60-93.

Hamacher, D.W. and Norris, R.P. (2010) Meteors in Australian Aboriginal Dreamings, *WGN – Journal of the International Meteor Organization*, 38(3), 87-98.

Hamacher, D.W. and Norris, R.P. (2011) Comets in Australian Aboriginal Astronomy, *Journal of Astronomical History & Heritage*, 14(1): 31-40.

Heffernan, J.A. (1984). *Papunya Luritja Language Notes*, Papunya: Papunya Literature Production Centre.

Hoogenraad, R. and Thornley, B. (2003) *The jukurrpa pocket book of Aboriginal Languages of Central Australia and the places where they are spoken*, Institute for Aboriginal Development (IAD) Press, Alice Springs

Kohman, T.P. and Goel, P.S. (1963) *Terrestrial ages of meteorites from cosmogenic 14C*. In "Radioactive dating", Proceedings of the Symposium on Radioactive Dating held by the International Atomic Energy Agency in co-operation with the Joint Commission on Applied Radioactivity (ICSU) in Athens, Greece on 19-23 November 1962, Volume 1962, p. 395-411

Madigan, C.T. (1937) The Boxhole Crater and the Huckitta Meteorite (Central Australia). *Royal Society of South Australia Transactions and Proceedings* 61:187-190

Madigan, C.T. and Alderman, A.R. (1939) The Huckitta Meteorite, Central Australia. *Mineralogical Magazine* 25(165): 353-371

Maegraith, B.G. (1932) The Astronomy of the Aranda and Luritja Tribes, *Transactions and proceedings and report of the Royal Society of South Australia* 56(1): 19-26

Malbunka, M. (2009) *Tnorala*, Message Stick, Aired on ABC-1 on Sunday 19 July 2009

Milton, D.J., Glikson, A.Y. and Brett, R. (1996) Gosse's Bluff - a latest Jurassic impact structure, central Australia. Part 1: geological structure, stratigraphy, and origin, *AGSO Journal of Australian Geology and Geophysics* 16(4): 453–486

Mitchell, J.M. (1934) Meteorite Craters - Old Prospector's Experiences. *The Advertiser*, Adelaide, South Australia, Thursday, 11 January 1934, p. 12

Mountford, C.P. (1976) *Nomads of the Australian Desert,* Rigby, Ltd., Adelaide

O'Connell, J.F. and Allen, J. (2004) Dating the colonization of Sahul (Pleistocene Australia–New Guinea): A review of recent research, *Journal of Archaeological Science* 31(6): 835-853

Parks and Wildlife Commission of the Northern Territory (1997) *Tnorala Conservation Reserve (Gosse's Bluff) Plan of Management*, March 1997, Alice Springs, Amended May 2007
http://nt.gov.au/nreta/parks/manage/plans/pdf/tnorala.pdf

Parks and Wildlife Commission of the Northern Territory (2002) *Henbury Meteorites Conservation Reserve: Draft plan of management*, November 2002
http://www.nt.gov.au/nreta/parks/manage/plans/pdf/henbury_pom.pdf

Róheim, G. (1945) *The Eternal Ones of the Dream: a psychoanalytic interpretation of Australian myth and *ritual*, International Universities Press, New York

Róheim, G. (1971) *Australian totemism: a study in psycho-analytic study in anthropology*, Frank Cass and Co, LTD, London

Sanday, P.R. (2007) *Aboriginal Paintings of the Wolfe Creek Crater: Track of the Rainbow Serpent*, University of Pennsylvania Press

Shoemaker, E.M.; Shoemaker, C.S.; Nishiizumi, K.; Kohl, C.P.; Arnold, J.R.; Klein, J.; Fink, D.; Middleton, R.; Kubik, P.W. and Sharma, P. (1990) Ages of Australian meteorite craters - a preliminary report. *Meteoritics* 25: 409

Shoemaker, E.M., MacDonald, F.A. and Shoemaker, C.S. (2005) Geology of five small Australian impact craters, *Australian Journal of Earth Sciences* 52: 529-544

Smith, M.A. (1987) Pleistocene occupation in arid Central Australia, *Nature* 328(6132): 710-711

Smith, M.A.; Bird, M.I.; Turney, C.S.M.; Fifiel, L.K.; Santos, G.M.; Hausladen, P.A. and di Tada, M.L. (2001) New abox AMS $^{14}$C ages remove dating anomalies Puritjarra rock shelter, *Australian Archaeology 53*: 45-47

Spencer, W.B. and Gillen, F.J. (1899) *The Native Tribes of Central Australia*, Reprinted 1968, Dover Publications, New York

Spencer, W.B. and Gillen, F.J. (1904) *The Northern Tribes of Central Australia*, Macmillan and Co., Ltd., New York

Spencer, W.B. and Gillen, F.J. (1927) *The Arunta: a study of a Stone Age people*, Macmillan and Co., Ltd., New York

Strehlow, C.F.T. (1907) *Die Aranda und Loritja-Stamme in Zentral Australien*, Joseph Baer and Co., Frankfurt, Germany

Thanguwa, T.S. (2008) "Kulu", http://www.thanguwa.com, accessed on 10 August 2010

Walsh, M. (1991) *Overview of Indigenous languages of Australia*, In *Language in Australia*, edited by S. Romaine, Cambridge University Press